\documentclass[conference]{IEEEtran}
\usepackage{amssymb}
\usepackage{graphicx}
\usepackage{algpseudocode}
\usepackage{algorithm}
\usepackage[utf8]{inputenc}
\usepackage[english]{babel}
\usepackage{csquotes}
\MakeOuterQuote{"}
\usepackage{xfrac}
\usepackage{mathrsfs}
\usepackage{yfonts}
\usepackage{amsmath}
\usepackage{tikz}
\usetikzlibrary{calc}
\usepackage{relsize}
\usepackage{xfrac}

\usepackage{lipsum}
\usepackage{graphicx}

\usepackage{amsmath}

\usepackage{amssymb, amsmath, amsthm, amsfonts}

\newtheorem{lemma}{Lemma}

\begin{document}


%

\long\def\/*#1*/{}

%

%
\ifCLASSINFOpdf
\else
\fi
\hyphenation{op-tical net-works semi-conduc-tor}

%
\title{ZeroBlock: Timestamp-Free Prevention of Block-Withholding Attack in Bitcoin}

\author{{Siamak Solat}
\IEEEauthorblockA{\\UPMC-CNRS, Sorbonne Universit\'es,\\
LIP6, UMR 7606, Paris, France\\
firstname.lastname@lip6.fr}
\and
{Maria Potop-Butucaru}

\IEEEauthorblockA{\\UPMC-CNRS, Sorbonne Universit\'es,\\
LIP6, UMR 7606, Paris, France\\
firstname.lastname@lip6.fr}}


%


\maketitle

\begin{abstract}
Bitcoin was recently introduced as a peer-to-peer electronic currency in order to facilitate transactions outside the traditional financial system.
The core of Bitcoin, the Blockchain, is the history of all transactions committed by the system. This distributed ledger is similar to a distributed shared register where miners write and read blocks.  New blocks in the Blockchain contain the last transactions in the system and are added by miners after a block mining process that consists in solving a difficult cryptographic puzzle.
Although, the reward is the main motivation for the mining process in Bitcoin, it also may be an incentive for attacks such as \emph{selfish mining}. 
In this paper we propose and theoretically analyze a solution for one of the major problems in Bitcoin : \emph{selfish mining} or \emph{block-withholding} attack. This attack is conducted by adversarial  miners in order to either earn undue rewards or waste the computational power of \emph{honest} miners. Contrary to the best to date solution for preventing \emph{block-withholding}, \cite{Heilman},  our solution, \emph{ZeroBlock}, prevents  this attack by using a novel timestamp-free technique that exploits the Poisson nature of the proof-of-work and the current knowledge on the propagation of information  in Bitcoin  \cite{Decker}. Note that previous solutions are vulnerable to forgeable time-stamps.
Additionally, our solution is compliant with miners churn.
\end{abstract}


%
\IEEEpeerreviewmaketitle

\section{Introduction}
In the last few years crypto-currencies \cite{Fugger,Saito,Vishnumurthy,Yang,Nakamoto,Bonneau,Miers} are in the center of the research ranging from financial, political and social to computer science and pure mathematics. Bitcoin \cite{Nakamoto} was one of the starters of this concentration of forces. It targeted the creation of a system where transactions between individuals can escape  the strict control of the banks and  financial markets. 

Bitcoin was introduced as a pure peer-to-peer  electronic currency or crypto-currency. It aims at fully decentralization of electronic transactions. Bitcoin allows to perform online  transactions directly from one party to another one "without" the interference of a financial institution as a "trusted third party" \cite{Nakamoto}. It uses digital signatures to verify the bitcoin ownership \footnote{Capitalized "Bitcoin" refers to the protocol, while lower case "bitcoin" refers to the coin.} and employs Blockchain in order to prevent double-spending attacks. In this attack the same bitcoin can be spent several times by a dishonest party.  Blocks in the blockchain are created via a proof-of-work (cryptographic puzzle)  \cite{Dwork,Back,Reid,ComprehensiveSurvey}
performed by \emph{honest} parties (miners that follow the protocol). 
Blockchain is further broadcasted via a peer-to-peer overlay in order to agree on a common  history of the transactions in the system. 

 Bitcoin is still vulnerable to various attacks including double-spending \cite{DSW16}, \emph{selfish} mining \cite{Eyal_Ittay}, Goldfinger \cite{Goldfinger}, 51\% attack \cite{Goldfinger} etc. In this paper we focus the \emph{selfish} mining attack.
Recently, \cite{Eyal} provided a  full description of incentives to withhold or \emph{selfish} mine in Bitcoin. That is, to force \emph{honest} miners to waste their computational power such that their public blocks become useless (as \textit{orphan} block), whereas the private chain of the \textit{selfish} miners is accepted as a part of the Blockchain. To this end, the \textit{selfish} miners reveal selectively their private blocks to make useless the blocks made by \emph{honest} miners. 

\emph{Our contribution.}
In this paper, we present and prove correct a new solution, ZeroBlock, which prevents \emph{block withholding}   or \emph{selfish} mining. ZeroBlock scheme, contrary to the recent solution proposed by \cite{Heilman}, does not use forgeable timestamps. Our solution builds on the following simple idea: if a \emph{selfish} miner keeps a block private more than a fixed interval of time, its block will be rejected by all the \emph{honest} miners. 
Zeroblock scheme strives  to reduce the probability of \emph{intentional} forks that are result of block-withholding attacks. 
With ZeroBlock scheme a selfish mining pool cannot achieve more than its expected reward. Only with a low probability, selfish mining pool may create intentionally an \emph{unprofitable} fork. We accentuate "unprofitable", because this fork does not lead to more reward for selfish mining pool, but also reduces selfish pool's likelihood to earn unexpected reward regardless of to its mining power. Thus, selfish mining pool is not incentivized to create such fork if its purpose is to achieve more reward. Furthermore, we prove that the maximum probability of such \emph{intentional} fork is very low ( $\approx 0.04$) when selfish pool uses its maximum hashing power. 
We further extend
ZeroBlock in order to be tolerant to miners churn.

\emph{Paper Roadmap.}
The rest of this paper is organized as follows: Section \ref{lbl:Bitcoin_Overview} presents an overview of  Bitcoin. Section \ref{lbl:Block_Withholding_or_selfish_mining} presents \emph{block withholding} attack or \emph{selfish} mining and briefly discuss the differences between an intentional and an accidental attack. 
 Section \ref{lbl:ZeroBlock} presents our ZeroBlock time-stamp free algorithm. 
 Section \ref{lbl:Eventualities}, proves that when ZeroBlock scheme is used a selfish mining pool cannot achieve more than its expected reward. Only with a low probability, selfish mining pool may create intentionally an \emph{unprofitable} fork.
 Section  \ref{lbl:ZeroBlockextension} extends ZeroBlock algorithm in order to make it compliant to miners churn. 
 Finally, 
Section \ref{lbl:Conclusion} concludes the paper.

\section{Bitcoin Overview}
\label{lbl:Bitcoin_Overview}
Bitcoin is an electronic coin which works as a chain of digital signatures where each owner transfers bitcoins to the next party after adding his signature (generated with his private key)  along with the hash of previous transactions and the public key of the receiver. As a result, the final receiver is able to verify the bitcoins ownership by verifying the signatures \cite{Nakamoto}. 

The key data structure in Bitcoin, called Blockchain, is a public log that is a sequence of blocks that maintain a linearized history of  transactions. The safety of Blockchain is ensured by a  cryptographic puzzle, named proof-of-work (PoW), solved by several nodes, named "miners". Miners who can solve PoW are permitted to generate a new block to record transactions  and  receive some bitcoins as a reward. This reward is used to further motivate miners to share their power resource with the network and continue to mine.

In Bitcoin system  builds on top of two key concepts. The first one are \emph{Transactions} : used to transfer coins in the network. The second one are \emph{Blocks}. Blocks are inserted into the \emph{public ledger} \emph{Blockchain}. Blockchains are further employed to synchronize  the state among all miners \cite{Decker} and to provide to each participant in the system a common view on the history of the transactions in the system. In the sequel we detail these two key notions.


\textbf{Transactions.} A \emph{transaction} transfers bitcoins from one or multiple source account addresses to destination account address. Each account address has a unique key. 
For transferring a bitcoin between two accounts, a \emph{transaction} is generated. This transaction includes the destination account address and it must be signed by the private key of the source account. 
To calculate the balance of the accounts, the public ledger calculates each \emph{transaction} output (i.e. a numeric value of coins).
Each \emph{transaction} is recognized by using the hash of its serialized context. 

When \emph{transactions} are broadcasted into the network, the public ledger will be updated locally by each miner. Without additional  care, this copy may be different at distinct miners which may lead to inconsistencies.  For example, consider that a miner receives a \emph{transaction} that shows the transfer of some bitcoins from some \emph{account "A"}, while the coins are not "available" for this particular account. In a different situation, several transactions may try to transfer the same coins more than once. Such subversive behavior is known as "\emph{double spending}" \cite{Decker}.    
In order to avoid that transactions appear into an inconsistent order at different miners in the network they are encapsulated in blocks that are further chained in a linearizable Blockchain.

\textbf{Blocks.} 
When a miner receives a transaction it starts a mining process in order to generate the block that will encapsulate the transaction.  Bitcoin network uses Hashcash \cite{Back} proof-of-work system for block generation such that a block is accepted by the network if miners perform proof-of-work properly and successfully. A proof-of-work (PoW) is a cryptographic puzzle that is difficult to solve but easy to verify.
The difficulty of PoW is adjustable regarding to the hashing power of the network. Currently, the block generation rate is set to one block per 10 minutes \cite{Bitcoin-NG}. 
Note that since the success of solving proof-of-work by a miner has very low probability, it is almost impossible to predict which miner or mining pool will generate the next block.

When a new block is generated, it is broadcasted in the entire network. If this new block is accepted as head of Blockchain, other miners start to work on this new block to extend the Blockchain. A competition between two new blocks may occur when their preceding block is equal and  they are broadcasted simultaneously. At this point Blockchain may fork. Since broadcasting a block takes only a few seconds but the average  time to discover a new block is around 10 minutes, thus,  \emph{accidental fork} occurs almost every 60 blocks \cite{Eyal,Decker}. In such situation, miners choose the first block which they receive and as a result, the second block will be ignored by the network as an \emph{orphan} block. Consequently, miners that worked on the second block wasted their computational power with no reward. 


\textbf{Blockchain and Forks in Bitcoin.} A \emph{Blockchain} is a chain of \emph{blocks} in "chronological order". Each \emph{block} has a \emph{parent block}. The \emph{genesis block} is the root. The most distant \emph{block} from \emph{genesis block} is called the \emph{head of Blockchain}. 

If two miners create two blocks with the same preceding block, Blockchain is \textit{forked} into two branches. This fork may occur \emph{accidentally} or \emph{intentionally}. 

Note that an \emph{accidental} fork occurs due to the nature and functionality of proof-of-work and it is not related to some particular attack as block-withholding or selfish mining.
Since proof-of-work is a Poisson process, two blocks may be discovered by two mining pools. The probability of an accidental fork is $\approx 1.69$ \cite{Decker} .  
 
In case of an \emph{intentional} fork, a \emph{selfish} miner generates and keeps private a block until it estimates opportunistic to reveal it.
When a  \emph{honest} miner generates a new block with the same preceding block as the  one generated by the  \emph{selfish} miner,  the latter propagates its private block to create an "intentional fork". The \emph{selfish} miner can generate  more than one private block in order to take the control of the Blockchain since the longest chain will be the only one commonly accepted by the honest miners in the network.  


\section{Block-Withholding Attack}
\label{lbl:Block_Withholding_or_selfish_mining}

\emph{Block withholding} attack was introduced as \emph{"Selfish mining attack"}  in \cite{Eyal} and also as \emph{"Block Discarding Attack"} in \cite{Bahack}. This attack relies on "block concealing" and revealing only at a special time selected by  \emph{selfish} miners or \textit{selfish} mining pool.  
According to \cite{Eyal}, these \emph{selfish} miners can earn revenues superior to  a fair situation \cite{Luu}.  
That is, the main purpose of block-withholding by selfish mining pool is achieving more rewards in comparison with its hashing power in the network. Thus, selfish mining pool's reward oversteps its mining power in the network and it can increase its expected mining reward.

In block-withholding strategy, a \emph{selfish} miner after solving proof-of-work and finding a new block does not broadcasts until a specific time. 

In \cite{Sapirshtein} authors extend the \emph{selfish} mining strategy and provide an algorithm to find optimal policies for \emph{selfish} miners. \cite{Kroll} shows that expanding the Blockchain by adding the newest block creates a simple model of "weak and non-unique" Nash equilibrium \cite{Sapirshtein}. \cite{Sapirshtein} shows that these optimal policies compute the threshold such that honest mining would be a "strict and  unique" Nash equilibrium. 

According to \cite{Eyal,Heilman}, the \emph{selfish} mining in Bitcoin network occurs as follows:
In case of generation of a new block by the \emph{honest} miners, (1) if the size of \emph{honest} branch is longer than the selfish branch, then the selfish cartel tries to set its private branch equal to the public branch. (2) If the selfish branch is one block more than the public branch, then selfish miners publish their private chain completely (3) If the selfish branch is more than one block longer than the public branch, then the selfish miners publish only the head of their private branch. In case of generation of a new block by selfish miners, they keep this new block private and in case of a competition with the \emph{honest} miners, they publish their private branch to win the competition.
According to \cite{Eyal}, the success of selfish miners in this competition is contingent on the parameters $\alpha$ (i.e. hashing power of selfish miners) and $\gamma$ (i.e. the hashing power of the \textit{honest} miners who work on the selfish branch). According to equation \ref{eq1}, if $\gamma = 0$, the threshold for success of block-withholding behavior is $\alpha$ $\ge$ 33 \% and if $\gamma = 0.99$, the threshold is $\alpha \ge 0.009$ \cite{Heilman}. \\ \\ 
\begin{equation}
	\frac{1 - \gamma}{3 - 2\gamma} < \alpha < \frac{1}{2}
\label{eq1}
\end{equation}

Eyal and Sirer \cite{Eyal} suggest a solution according to which $\gamma$ is fixed to 0.5 and consequently the threshold of successful block-withholding is $\alpha$ $\ge$ 0.25. 
Heilman \cite{Heilman} introduces an approach named "Freshness Preferred" (FP). Using  random beacons and timestamps, \emph{honest} miners select more fresh blocks and the threshold becomes $\alpha$ $\ge$ 0.32.

The rest of the paper is organized as follows. In Section \ref{lbl:ZeroBlock} we detail  our ZeroBlock scheme that targets to reduce the probability of \emph{intentional} forks that may result after block-withholding attacks appear in the network. 
In Section \ref{lbl:Eventualities} we prove that using ZeroBlock idea a selfish mining pool cannot achieve more than its expected reward. Only with a low probability, selfish mining pool may create intentionally an \emph{unprofitable} fork. We accentuate "unprofitable", because this fork does not lead to more reward for selfish mining pool. Thus, selfish mining pool is not incentivized to create such fork. Furthermore, we prove that the \emph{maximum} probability of such \emph{intentional} fork is very low (maximum $\approx 0.04$) when selfish pool uses its maximum hashing power. Furthermore, we extend the ZeroBlock in order to make it resilient to miners churn.


\section{ZeroBlock Algorithm}
\label{lbl:ZeroBlock}
In this section we introduce a new solution, ZeroBlock (see Algorithm \ref{euclid}) 
to prevent \emph{block-withholding}  or \emph{selfish} mining in Bitcoin. 
The key idea of our solution is that each block must be generated and received by the network within \emph{a maximum acceptable time for receiving a new block} interval, \textit{mat} (see equation \ref{eq7} below). Within a $mat$ interval a \emph{honest} miner receives or discovers a new block. Otherwise, it generates a dummy block. The computation of each $mat$ interval is done locally by each miner based on the following Bitcoin parameters: the \emph{expected delay for a block mining} and the \emph{information propagation time} in the Bitcoin network. The former parameter is discussed in details below while the latter has been extensively studied in \cite{Decker} where the authors shown that a published  block is received by the whole network within 60 seconds (see Figure \ref{fig:IPT} in the appendix).
 
 \emph{Expected delay for a block mining} in Bitcoin depends  mainly on the difficulty of proof-of-work. The major part of proof-of-work consists in discovering a byte string, \textit{nonce}.
 As pointed out in  \cite{Decker} proof-of-work in Bitcoin is a Poisson process and causes blocks to be discovered randomly and independently. Moreover,  as advocated in \cite{Bitcoin-NG,Bitcoin mining,Bitcoin bespoke,Bitcoin Difficulty,Bitcoin Trends,Bitcoin empirical analysis,Bitcoin architectural analysis,Bitcoin Miner Evolution}
  in Bitcoin, the difficulty of proof-of-work required to discover a block is periodically adjusted such that, on average, \textit{one} block is expected to be discovered every \textit{10 minutes}. 
Hence,  the difficulty of proof-of-work is updated every 2016 blocks. It means that regarding to this adjustment (i.e. one block per 10 minutes) 2016 blocks, on average, is expected to be generated in 14 days. If 2016 blocks are discovered in a shorter time, the difficulty of proof-of-work will be increased and if they are generated in a longer time, difficulty of proof-of-work will be decreased. 


In more details, miners for each input of proof-of-work (i.e. a random \textit{nonce}) calculate a hash value. This hash is a number between 0 and a maximum value of a 256-bit number. The miner has discovered the answer of proof-of-work, if and only if this hash is below the \textit{target}. 

The proof-of-work works as follows:\\
\begin{equation}
	i\!f\hspace{0.10cm} H(pb + nonce) < T\hspace{0.10cm} then
\label{eq2}
\end{equation}
\begin{center}
	\textit{proof-of-work succeeded}
\end{center} 
\vspace{5mm}

where \textit{pb} is representing the hash of the previous block, \textit{nonce} is the answer of proof-of-work that must be found by miners , \textit{T} is \textit{target}, '+' is concatenation operation and \textit{H} is the hash function.

Each mining pool can estimate the difficulty of proof-of-work using equation \ref{eq3}.

\begin{equation}
	D = \frac{maxTarget}{T}
\label{eq3}
\end{equation} 

where \textit{D} is the difficulty of proof-of-work, \textit{T} is current \textit{target} and \textit{maxTarget} is maximum possible value for \textit{target} that is ($2^{16}$ - 1)$2^{208}$ $\approx 2^{224}$. Since the hash function produces uniformly
a random value between 0 and $2^{256} - 1$ thus, the probability that a given \textit{nonce} value would be the answer of proof-of-work is as follows (equation \ref{eq4}):

\begin{multline}
Prob(\textit{nonce is answer}) =  \\
\dfrac{target}{2^{256}} = \dfrac{2^{224}}{D \times 2^{256}} \approx \dfrac{1}{D \times 2^{32}}
\label{eq4}
\end{multline}

The number of hashes to discover a block is \textit{D} $\times$ $2^{32}$  in expectation. If a mining pool  can calculate hashes at a rate \textit{php} (we call this as pool's hashing power), then the expected time (or average time) \textit{avt} in which this pool can discover a block is as follows (equation \ref{eq5}):

\begin{equation}
	avt_{pool} = \dfrac{D \times 2^{32}}{php}
\label{eq5}
\end{equation}

When we replace \textit{php} by hashing power of the network, \textit{nethp}, we can use equation \ref{eq4} for the entire network as follows (equation \ref{eq6}):

\begin{equation}
	avt_{net} = \dfrac{D \times 2^{32}}{nethp}
\label{eq6}
\end{equation}
 
According to the relation between \textit{time, difficulty of proof-of-work, hashing power of the network} in equation \ref{eq6}, Bitcoin network adjusts \textit{D} such that regarding to hashing power of the network, the average time for block generation rate remains 10 minutes.

To calculate  the \textit{the maximum acceptable time for receiving a new block}, \textit{mat}, we  use equation \ref{eq7} below:

\begin{equation}
	mat = avt_{net} + ipt
\label{eq7}
\end{equation}

where $avt_{net}$ is given by the equation \ref{eq6} and $ipt$ is the information propagation time in Bitcoin network as estimated in \cite{Decker}.




\subsection{Zeroblock Algorithm parameters and notations}
\label{lbl:Key notations}
The ZeroBlock algorithm (Algorithm \ref{euclid}) uses the following parameters and definitions:
\begin{itemize} 
\item \textit{ipt} : information propagation time in Bitcoin network that is an average delay for propagation a block into the network. This average delay has been estimated by simulation in \cite{Decker} (see also Figure 1 in the Appendix). 
\item  \textit{avt} : block generation rate that has been set by Bitcoin protocol according to which the difficulty of proof-of-work is adjusted regarding to the hashing power of the network using equation \ref{eq6}. 
\item  \textit{mat} : maximum acceptable time for receiving a new block that is computed by equation \ref{eq7}.  During a $mat$ interval if a miner cannot solve the proof-of-work, it has to generate a dummy Zeroblock.
\item \textit{unpermitted block-withholding} : occurs when a \textit{selfish} mining pool discovers a new block and keeps the block private after the end of the current $mat$ interval. 
\item \textit{Dummy Zeroblock} : is generated locally by miners. It includes the index of  $mat$ interval and the hash of previous block. It is generated by honest miners to prevent \emph{unpermitted block-withholding}. Note that our solution uses  \emph{standard Bitcoin blocks} discovered by solving the proof-of-work and \emph{dummy blocks} that are generated by the  Zeroblock algorithm for which miners do not need to solve any proof-of-work.  The dummy Zeroblocks time  generation is therefore  ignored when adjusting the difficulty of the proof-of-work. 
\item  \textit{orphan block} : a block that has been discovered but is then rejected by the network. 
\item \textit{genesis block} : the first block of a Blockchain on which all miners have a consensus. 
\item \textit{correct chain} : a chain whose blocks have been discovered and inserted correctly according to the described protocol. 
\item \textit{creative miner} : a miner that in a $mat$ interval can solve proof-of-work and then generates a new block. 
	
	
\end{itemize}

\begin{algorithm*}
\caption{ZeroBlock algorithm}\label{euclid}
\begin{algorithmic}[1]
\State $\textit{index} \gets 0$ \Comment{index of \textit{mat}} \label{lst:line:1}
\State $\textit{mat[index]} \gets 0$ \Comment{\textit{mat} at the beginning is set to zero} \label{lst:line:2}
\State $avt_{net} \gets \textit{block generation average time}$ \Comment{according to equation (6)} \label{lst:line:3}
\State $\textit{localChain} \gets \textit{Genesis}$ \label{lst:line:4}
\State $\textit{FlagNewBlock} \gets \textit{False}$ \label{lst:line:5}
\State $\textit{nonce} \gets 0$ \label{lst:line:6}
\State $\textit{HPrB} \gets 0$ \Comment{hash of previous block} \label{lst:line:7}
\State $\textit{T} \gets \textit{target}$ \label{lst:line:8}
\State $\textit{newChain} \gets \textit{Null}$ \label{lst:line:9}
\State $\textit{ansPoW} \gets 0$ \Comment{answer of PoW} \label{lst:line:10}
\State $\textit{scounter()} \gets 0$ \Comment{scounter() is a seconds counter} \label{lst:line:11}
\While{($\textit{True}$)} \label{lst:line:12}
    \If{($\textit{FlagNewBlock = False)}$ $AN\!D$ ($\textit{mat[index]}$ $\not=$ 0)} \label{lst:line:13}
    	\State $\textit{dummy Zeroblock} \gets \textit{SHF(getHead(localChain)) + SHF("FixedStringZB") + index}$\label{lst:line:14}  
    	\State $\textit{localChain} \gets \textit{join(dummy Zeroblock,localChain)}$ \label{lst:line:15}
    \EndIf \label{lst:line:16}
    \State $\textit{index} \gets \textit{index + 1}$ \label{lst:line:17} 
    \State $\textit{refresh(mat[index])}$ \label{lst:line:18}
    \While{($\textit{scounter()}$ $\le$ $\textit{mat[index]}$)} \label{lst:line:19}
    	\State $\textit{newChain} \gets \textit{checkInput()}$ \label{lst:line:20}
    	\If{($\textit{newChain}$  $\not=$  $\textit{Null}$)} \label{lst:line:21}
    		\State $\textit{HPrB} \gets \textit{SHF(getHead(localChain))}$ \label{lst:line:22}
    		\If{($\textit{FHF(HPrB,newChain.ansPoW)}$ $\le$ $T$)}\label{lst:line:23}   \Comment{proof-of-work is done}         
    			\State $\textit{localChain} \gets \textit{newChain}$ \label{lst:line:24}
    			\State $\textit{newChain} \gets \textit{Null}$ \label{lst:line:25}
    			\State $\textit{FlagNewBlock} \gets \textit{True}$ \label{lst:line:26}
    			\State $\textit{Break}$ \label{lst:line:27}
    		\EndIf \label{lst:line:28}
    	\EndIf \label{lst:line:29}
    	\If{($\textit{scounter()}$ $< avt_{net}$)} \label{lst:line:30} 
    	\If{($\textit{FlagNewBlock = False}$)} \label{lst:line:31} 
    			\State $\textit{HPrB} \gets \textit{SHF(getHead(localChain))}$ \label{lst:line:32}
    			\If{($\textit{FHF(HPrB , nonce)}$ $\le$ $T$)} \Comment{proof-of-work succeeded} \label{lst:line:33}
    				\State $\textit{ansPoW} \gets \textit{nonce}$ \label{lst:line:34}
    				\State $\textit{localChain} \gets \textit{join(GenerateBlock(),localChain)}$ \label{lst:line:35}
    				\State $\textit{BroadcastBlock(localChain,ansPoW)}$ \label{lst:line:36}
    				\State $\textit{FlagNewBlock} \gets \textit{True}$ \label{lst:line:37}
    				\State $\textit{nonce} \gets 0$ \label{lst:line:38}
    				\State $\textit{Break}$ \label{lst:line:39}
    			\EndIf \label{lst:line:40}
    			\State $\textit{nonce} \gets \textit{nonce + 1}$ \label{lst:line:41}
    	\EndIf \label{lst:line:42}
    	\EndIf \label{lst:line:43}
	\EndWhile \label{lst:line:44}
\EndWhile \label{lst:line:ZB_45}

\end{algorithmic}
\end{algorithm*}

\subsection{ZeroBlock Algorithm Detailed Description}
\label{lbl:ZeroBlock Algorithm}
In this section, we describe the ZeroBlock algorithm shown as Algorithm \ref{euclid}. 
Each miner, $\mu$, that executes Algorithm \ref{euclid}  performs the following steps: 
\begin{itemize}
\item (\emph{lines} \ref{lst:line:1} to \ref{lst:line:11}) initialization: where $mat_{0}$ and \emph{scounter()} (seconds counter) are set to 0. 
\item (\emph{line} \ref{lst:line:4}) miner $ \mu $ initiates its local chain by Genesis block. 
\item (\emph{line} \ref{lst:line:5}) \textit{FlagNewBlock} is set to \textit{False}. 
\item In (\emph{line} \ref{lst:line:12}) miner $ \mu $ starts an infinite loop. 
\item In (\emph{line} \ref{lst:line:13}) miner $ \mu $ checks (\textit{FlagNewBlock} = \textit{False}) (that means it verifies if there is no new block) and also ($mat_{index} \neq 0$). For the first time, $mat_{0} = 0$ and thus the second condition is not satisfied. 
\item In (\emph{line} ~\ref{lst:line:17}) miner $ \mu $ increases \textit{index} and then in  (\emph{line} \ref{lst:line:18}) invokes \textit{refresh()} function that performs equation \ref{eq8}.  

\begin{equation}
mat_{index} = mat_{index - 1} + (avt_{net} + ipt)
\label{eq8}
\end{equation}

\item (\emph{line} \ref{lst:line:19}) While miner $ \mu $ has time to discover and broadcast a new block (\emph{scounter() $\le mat_{index}$}) in (\emph{line} \ref{lst:line:20}) checks its input to know if there is a new block received from the network. 
\item (\emph{line} \ref{lst:line:21}) If yes, in (\emph{line} \ref{lst:line:23}) miner $ \mu $ verifies the answer of PoW for the new block. 
\item (\emph{line} \ref{lst:line:24}) If PoW has been done correctly, $ \mu $ replaces the local chain by the new chain and in (\emph{line} \ref{lst:line:26}) updates the value of \textit{FlagNewBlock = True} and in (\emph{line} \ref{lst:line:27}) leaves the while loop and goes to (\emph{line} \ref{lst:line:13}) and then since \textit{FlagNewBlock = True} goes to (\emph{line} \ref{lst:line:17}). 
\item If there is no new block in its input and \textit{scounter()} is below $avt_{net}$, then miner $ \mu $ tries to solve PoW in (\emph{line} \ref{lst:line:33}). 
\item If miner can solve PoW, then it generates a new block in (\emph{line} \ref{lst:line:35}) and broadcasts it in (\emph{line} \ref{lst:line:36}). 
\item In case the given \textit{nonce} is not the answer of PoW (the condition of (\emph{line} \ref{lst:line:33}) is incorrect), miner $ \mu $ increases the \textit{nonce} (in \emph{line} \ref{lst:line:41}) and goes back to (\emph{line} \ref{lst:line:19}). 
\item If \emph{scounter()} is more than $mat_{index}$, immediately miner $ \mu $ generates a dummy Zeroblock in (\emph{line} \ref{lst:line:14}) and then adds the dummy Zeroblock to its local chain (\emph{line} \ref{lst:line:15}). \\ 
\item Then, miner $ \mu $ refreshes index and value of \emph{mat} in (\emph{lines} \ref{lst:line:17} and \ref{lst:line:18}).  
\item If a dummy Zeroblock is generated, miner $ \mu $ rejects a \emph{selfish} block in (\emph{line} \ref{lst:line:23}) because the answer of proof-of-work for selfish block is not the correct \textit{nonce}, since it does not include the hash of the dummy Zeroblock. Then miner $ \mu $ goes to (\emph{line} \ref{lst:line:30}) and repeats the algorithm as described above. 
\end{itemize}


\section{ZeroBlock Algorithm resilience to  block-withholding attack}
\label{lbl:Eventualities}      
First we prove that honest miners never accept chains infected with unpermitted block withholding. Then we prove that selfish miners are not incentivized to withhold blocks or not follows the Zeroblock algorithm.

\begin{lemma}
\label{th:1}
Honest miners, regardless of their percentage in the network, never accept chains infected with unpermitted block-withholding.
\end{lemma}

\begin{figure*}
\centering
\makebox[\textwidth][c]{\includegraphics[width=0.5\textwidth]{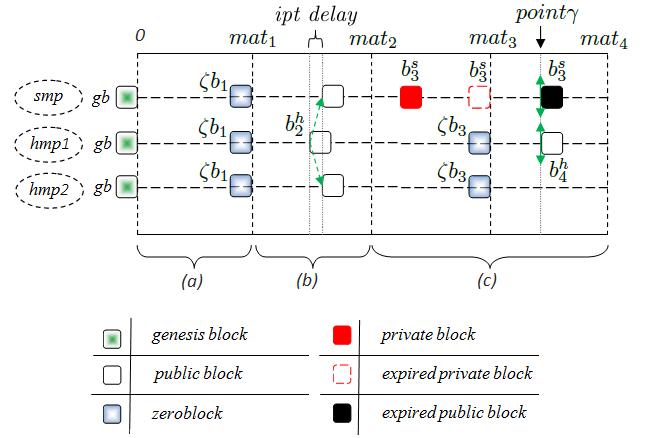}}
\caption{\textit{smp}: selfish mining pool, $hmp_{1}$: first honest mining pool, $hmp_{2}$: second honest mining pool. \textit{Part (a)} represents \textit{event 1}: In $mat_{1}$, neither honest nor selfish pools discover a new block. \textit{Part (b)} represents \textit{event 2}: In $mat_{2}$, the first pool which discovers a new block is the honest mining pool. \textit{Part (c)} represents \textit{event 3}: In $mat_{3}$, the first pool which discovers a new block is selfish mining pool, then at point $\gamma$ when honest pool discovers block $b_{4}^h$ selfish pool broadcasts $b_{3}^s$ to create an intentional fork, but thanks to dummy Zeroblock $\zeta b_{3}$ it will be rejected by honest pools since it does not include the hash of dummy Zeroblock $\zeta b_{3}$.}
\label{fig:Event_1_2_3(5)}
\end{figure*}

\begin{proof} 
In a $mat_{i}$ interval, a \emph{honest} miner either generates or receives a new block $b_{i}$, otherwise it generates a dummy Zeroblock $\zeta b_{i}$. 

An \emph{unpermitted block-withholding} occurs if a \emph{selfish} miner discovers a new block $b^s_{i}$ but keeps the block private after end of $mat_{i}$. At this point, to prevent \emph{unpermitted block-withholding} , \emph{honest} miners generate a dummy Zeroblock $\zeta b_{i}$ including the hash of previous block, since they have not received new block $b^s_{i}$ which is discovered by a \textit{selfish} miner in $mat_{i}$ interval. 

When $mat_{i}$ interval is finished (\emph{line} \ref{lst:line:19}), \emph{honest} miners leave the \textit{while} loop and go to the (\emph{line} \ref{lst:line:12}). Since two conditions in (\emph{line} \ref{lst:line:12}) are satisfied, they generate a Zeroblock $\zeta b_{i}$. This is due to the fact that a new block $b_{i}$ has been received (thus \textit{FlagNewBlock = False}) and $mat_{i} \not= 0$ (because, in (\emph{line} \ref{lst:line:12}), $ mat_{i} $ has been updated according to equation \ref{eq8}). A \emph{selfish} miner (which here is a \emph{creative miner}) has kept a new block $b^s_{i}$ after the end of $mat_{i}$ (see Figure \ref{fig:Event_1_2_3(5)} that shows a similar situation in which $ b^s_{3} $ that is discovered in $ mat_{3} $ by a selfish mining pool, \textit{smp}, has been kept private after the end of $ mat_{3} $). Thus, the block $b^s_{i}$ will not be accepted by \emph{honest} miners, because they have generated a dummy Zeroblock $\zeta b_{i}$ at the end of $mat_{i}$. Therefore, proof-of-work must be restarted from beginning for discovering the next block, i.e. $b_{i+1}$, such that its proof-of-work includes the recent $\zeta b_{i}$. At this point, $b^s_{i}$ is not acceptable since its proof-of-work has not been solved based on $\zeta b_{i}$.  

In other words, 
\begin{center}
	$\forall\ mat_{\Delta } :$ \\ 
	$\qquad\qquad\qquad\textit{ If } (\Delta \neq 0) \textit{ then }$ \\ 
	$\quad\qquad\qquad\qquad\qquad\qquad\qquad\exists\ (b_{\delta} \ \lor \ \zeta b_{\delta}) , \delta \geq \Delta$ \\ 
	$\quad\textit{ else }$ \\ 
	$\qquad\qquad\qquad\exists\ gb$ \\ 
\end{center}

Thus,

\begin{center}
	$\forall\ mat_{\Delta } :$ \\ 
	$\qquad\qquad\qquad\qquad b_{\delta} \in \textit{X , If } (\delta < \Delta)$ \\ 
\end{center}

where $\Delta$ is the index of a $mat $, $ \delta $ is the index of blocks, $X $ is the set of \textit{selfish} blocks, $b$ is a standard blocks, $ \zeta b$ is a dummy Zeroblock, and $ gb $ is the \textit{genesis} block. As shown in Figure \ref{fig:Event_1_2_3(5)}, where block $ b^s_{3} $ belongs to selfish mining pool in $mat_{4}$, it is rejected by honest miners, since $\delta < \Delta$, where $\delta$ = 3 and $\Delta$ = 4. 

\end{proof}

\begin{figure*}
\centering
\makebox[\textwidth][c]{\includegraphics[width=0.3\textwidth]{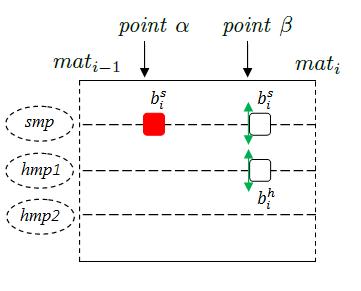}}
\caption{The selfish pool, \textit{smp},  discovers a new block $b^s_{i}$ at the point $ \alpha $ and keeps the block privately till point $ \beta $ when an honest pool, $hmp_{1}$, generates block $b^h_{i}$. At this point, selfish pool broadcasts $b^s_{i}$ to create an intentional fork. The maximum probability of this event  is ($\approx \mathit{0.04}$). 
}
\label{fig:Event_4(3)}
\end{figure*}

In the following, we prove that selfish miners are not incentivized to withhold blocks or to not follow correctly the dummy Zeroblocks generation.
In the following, we describe all possible events in a  \textit{maximum acceptable time for receiving a new block} interval. To simplify, we consider that the network consists of two mining pool types, including: two honest mining pools ($hmp_{1}$ and $ hmp_{2} $) and a selfish mining pool ($ smp $). We also assume that at time $t = 0$ all mining pools have a consensus on the first block, \textit{genesis} block. 

\paragraph{Event 1 - In a $ mat_{i} $ interval, neither honest nor selfish pools discover a new block} In this situation, a dummy Zeroblock $\zeta b_{i}$ will be generated at the end of $ mat_{i} $ interval by all honest pools. We recall that dummy Zeroblock generation time is negligible. Consequently, the hash of this dummy Zeroblock will be used for discovering the next block. It means that the answer of proof-of-work (\textit{nonce}) for discovering the next block depends on this dummy Zeroblock (since hash of previous blocks is used in proof-of-work and thus affects its answer). As a result, if a selfish mining pool does not generate this dummy Zeroblock, then it will not be able to find the correct answer of proof-of-work for next block (since all honest mining pools at time of verifying answer of proof-of-work will reject any \textit{nonce} that has not been earned from hash of this dummy Zeroblock.) (see Figure \ref{fig:Event_1_2_3(5)}, \textit{Part (a)}.)

\paragraph{Event 2 - In a $ mat_{i} $, the first pool which discovers a new block is the honest mining pool} In this case, it immediately broadcasts the discovered block to the entire network and then starts to discover the next block. Then, other pools receive this new block within \textit{ipt} time interval (i.e. \textit{information propagation time in Bitcoin network} which is simulated and estimated in \cite{Decker}.) Thus, other mining pools after receiving and verifying this new block starts to discover  the next block. As a result, the block creator will begin to discover  the next block a little sooner than the rest of the network. The maximum of this time is \textit{ipt}. 
 This might be considered as a time reward for the block creator that increases the miners' motivation to be the first one who discovers a new block. This time advantage dedicated to block creator is not unfair since creator mining pool is the first one who discovered the new block and for this it is fair to receive a time reward. However, this time difference according to simulations in \cite{Decker} is less than one minute (see Figure \ref{fig:Event_1_2_3(5)}, \textit{Part (b)}).

\paragraph{Event 3 - In a $ mat_{i} $, the first pool which discovers a new block is selfish mining pool} In this case, selfish pool keeps the block private until the end of $ mat_{i} $ interval and does not broadcast its block. We assume that by this time, honest mining pools could not discover a new block. Thus, honest pools generate a dummy Zeroblock $ \zeta b_{i} $ and restart to find the answer of proof-of-work based on hash of $ \zeta b_{i} $. Consequently, the next new block is acceptable by honest pools if its proof-of-work has been solved base on $ \zeta b_{i} $. As a result, this selfish block will be rejected by honest pools. (see Figure \ref{fig:Event_1_2_3(5)}, \textit{Part (c)}, \textit{point $\gamma$}.) 

\paragraph{Event 4 - In a $ mat_{i} $, the first pool which discovers a new block is the selfish mining pool ($b_i^s$ at the point $ \alpha $ in Figure \ref{fig:Event_4(3)})} In this case, selfish pool decides to keep the block private until the end of $ mat_{i} $. At the point $ \beta $ (see Figure \ref{fig:Event_4(3)}) an honest mining pool discovers a new block, $ b^s_{i} $. As soon as honest pool broadcasts the block $b_i^h$, selfish pool broadcasts block $b_i^s$ to create an \textbf{accidental} fork. Thus, a part of network receives $b_i^s$ and works on this block and another part works on $b_i^h$. As a result, with a probability one of $b_i^s$ or $b_i^h$ will be winner block and another one is ignored as \textit{orphan} block and eventually the winner block creator ($ smp $ or $ hmp_1 $) will receive the respective reward. Whereas, if selfish pool at point $ \alpha $ had broadcast its block, $b_i^s$, it had received the reward as the first block creator without any rival with probability of 100\%. Consequently, selfish pool with delay in broadcasting its block, $b_i^s$, caused to reducing probability of earning the reward. An action which is in contrast to the main purpose of block-withholding that is achieving more reward. We called this event as \textit{unprofitable block-withholding}. As a result, the selfish mining pool is not incentivized to reduce its chance for receiving the reward. 

In the following
we calculate the \textit{maximum} probability of \textit{event 4}, when the selfish mining pool has maximum possible percentage of hashing power of the network. Regarding to \textit{51\% attack}, the selfish pool  can have at most $49\%$ of total hashing power of the network, because otherwise selfish mining pool can get control of the network. 

Recall that the proof-of-work is a Poisson process and also the difficulty of proof-of-work is adjusted regarding to hashing power of the network such that the expected time to discover the next block is 10 minutes. Thus, the probability that $ \rho $ blocks to be discovered in a time interval in which we expect the network discovers $\lambda$ blocks, is given by equation \ref{eq8}.

\begin{equation}
	Prob(\rho|\lambda) = \dfrac{e^{-\lambda} . \lambda ^ \rho}{\rho!}
\label{eq9}
\end{equation}    

For example, the probability of discovering one, two, three, four and five blocks in 10 minutes, in descending order, is as follows:

\[
Prob(\rho=1|1) = e^{-1} \approx 0.3679
\]
\[
Prob(\rho=2|1) = \frac{e^{-1}}{2!} \approx 0.1839
\]
\[
Prob(\rho=3|1) = \frac{e^{-1}}{3!} \approx 0.0613
\]
\[
Prob(\rho=4|1) = \frac{e^{-1}}{4!} \approx 0.0153
\]
\[
Prob(\rho=5|1) = \frac{e^{-1}}{5!} \approx 0
\]
\begin{flushleft}
where: the expected number of blocks to be discovered in 10 minutes is $\lambda$ = 1.
\end{flushleft}

The maximum probability that in 10 minutes two blocks are discovered such that the first one is generated by selfish mining pool with the hashing power of $\mathit{sp = 0.49}$ and the second one is generated by honest mining pool with the hashing power of $\mathit{hp = 0.51}$ is as follows:

\begin{center}
$
\mathit{\textit{Maximum Possible Probability }(\textit{Event 4 } ) =} 
$
\begin{center}
$
\mathit{ = \big(sp . e ^ {-sp}\big) \big(hp . e ^ {-hp}\big) \times{sp}}
$
\end{center}
\end{center}
\[
\mathit{\approx 0.3 \times{0.3} \times{0.49}  \approx 0.04}
\]

Note that as we mentioned above if \textit{event 4} occurs, selfish pool reduces its likelihood for receiving the respective reward. 
Even if the purpose of selfish pool is only the network sabotage, it has to deplete its limited computational resource for achieving an event that in the \emph{best case for the selfish pool} has a low probability (at most $\approx \mathit{0.04}$). This leads to reduce selfish pool's motivation to perform such behavior. Note that this probability is for the case the selfish  pool has its \emph{maximum possible} hashing power (i.e. 49\% of total hashing power of the network) regarding to 51\% attack.

\section{Zeroblock Algorithm extension to dynamic miners}
 \label{lbl:ZeroBlockextension}
In the following we show how Algorithm \ref{euclid} can be extended to environments where miners can join at any time. First, we describe a simple mechanism that assumes that a new miner that joins the system has the ability to communicate with all the other miners in the network. 
When a new miner joins the network it broadcasts a message including its address to announce its entrance. The other miners respond with the last version of their local chain. Then, the new miner compares the received chains and selects the chain  belonging to the majority. Thus, if half plus one miners of the network are \emph{honest}, the \emph{honest} chain will be chosen by the new miner and it will further mine the \emph{correct chain}. Then, the new miner, similar to other miners, will perform the protocol as described in Algorithm \ref{euclid}. According to this process, each miner is able to leave and re-join the network infinitely.




Note that the above describe process uses as  hypothesis  that a new miner connects to the entire network. However, in current Bitcoin protocol a new node connects to only a subset of nodes in the network (usually 8 randomly selected nodes). If we assume a situation in which a new node connects to a subset of nodes, then the new node achieves the \textit{correct chain} with probability $P$ as follows: 


\begin{center}  
$
\textrm{P}(h \geq [\sfrac{\sigma}{2}] + 1) = \displaystyle \sum_{i=1}^{[\sfrac{\sigma}{2}]} \textrm{P}(h = [\sfrac{\sigma}{2}] + i)
$
\end{center} 
\begin{center}  
$
= \displaystyle \sum_{i=1}^{[\sfrac{\sigma}{2}]} \dfrac{\binom \eta {[\sfrac{\sigma}{2}] + i }\binom \psi {\sigma - ([\sfrac{\sigma}{2}] + i)}}{\binom n \sigma}
\vspace{10px}
$
\end{center}

where $n$ is the network size, $\sigma$ is the number of selected nodes to connect, $\eta$ is the number of $\textit{honest}$ nodes in the entire network, $\psi$ is the number of $\textit{selfish}$ nodes in the entire network and \textit{h} is the number of $\textit{honest}$ nodes in the set of selected nodes.

Table \ref{Table Prob} shows some numerical examples in different situations, each of which includes $\eta$ \textit{honest} nods and $\psi$ \textit{selfish} (\textit{adversarial}) nodes. We  use practical values for the size of the network, \textit{n}, and the number of selected nodes for connection, $\sigma$. We thus set $n$ and $\sigma$ to 5000 \cite{BitcoinSize} and 8 \cite{Decker}, respectively.  

\begin{table}
	\centering
	\begin{tabular}{| c | c | c | c | c |} 
		\hline
		$ \eta $ & $ \eta $ percentage & $ \psi $ & $ \psi $ percentage & P \\ [0.5ex] 
		\hline\hline
		4750 & 95\% & 250 & 5\% & $ \approx $ 0.9994\% \\ 
		\hline
		4250 & 85\% & 750 & 15\% & $ \approx $ 0.9785\% \\
		\hline
		3750 & 75\% & 1250 & 25\% & $ \approx $ 0.8862\% \\
		\hline
		3250 & 65\% & 1750 & 35\% & $ \approx $ 0.7063\% \\ 
		\hline 
	\end{tabular}
	\caption{The probability that at least half plus one nodes from the set of selected nods, $ \sigma $ are honest. In each example, the network includes $\eta$ \textit{honest} nods and $\psi$ \textit{selfish} (\textit{adversarial}) nodes. We use practical values for the size of the network, \textit{n}, and the number of selected nodes for connection, $\sigma$. We thus set $n$ and $\sigma$ to 5000 \cite{BitcoinSize} and 8 \cite{Decker}, respectively.}
	\label{Table Prob}
\end{table}

In the following,
we say two chains $ C_{1} $ and $ C_{2} $ are \emph{homogeneous}, if and only if both of them belong to the \emph{honest} pools or both of them belong to the \emph{adversarial} pools. Otherwise chains are called \emph{inhomogeneous}.


Note that 
since all honest chains are equal to each other, thus the new node is able to distinguish a set of \textit{homogeneous} chains. We propose if a new node receives a set of \textit{inhomogeneous} nodes, then it must \emph{retry} to connect to $ \sigma $ nodes till the new node receives a set of \textit{homogeneous} nodes. In this case, the new node eventually achieves a set of chains that belong to the adversary  if and only if all $ \sigma $ chains are  not correct chain. Thus, the adversarial cartels size needs to be increased significantly such that with a good probability all $ \sigma $ chains would not be correct chains.  Assume a situation in which  only one chain is correct chain and the rest are not. Then, the new node will retry to connect again to $ \sigma $ nodes, whilst if the new node consider the majority chains, then it will accept the adversarial chain.

Thus, we calculate the probability that  all $ \sigma $ chains are homogeneous and  \emph{correct} and then the probability that all $ \sigma $ chains are homogeneous and are \emph{not correct chain}, respectively, as follows: 

\begin{center}
$
\textrm{P}_{(hcorr)} = \textrm{P} (\textit{h = $ \sigma $}) = \dfrac{\binom \eta \sigma}{\binom n \sigma}   
$
\end{center}

\begin{center}
$
\textrm{P}_{(hnotc)} = \textrm{P} (\textit{h = 0}) = \dfrac{\binom \psi \sigma}{\binom n \sigma}
$
\end{center}

and then, we calculate the probability that all $ \sigma $ chains are homogeneous regardless of chains type as follows:

\begin{center}
	$
	\textrm{P}_{(hom)} = \textrm{P} (\textit{h = $ \sigma $ or h = 0}) = \dfrac{\binom \eta \sigma + \binom \psi \sigma}{\binom n \sigma} 
	$
\end{center}

We define R as the number of trials for connection to $ \sigma $ nodes to achieve a set of homogeneous chains. Thus, the probability that after \textit{m} trials \footnote{When \textit{m} = 1 it means that the new  tried to connect to $ \sigma $ nodes previously but it has not achieved a set of \textit{homogeneous} chains.} the new node achieves a set of \textit{homogeneous} and \textit{correct chains} is as follows:

\begin{center}
	$
	\textrm{P} (\textrm{R} = \textit{m} \ | \ \forall c_{i} \in \textrm{SC} , c_{i} \ \textit{is correct chain}) = 
	$
\end{center}

\begin{center}
$\bigg(1 - \textrm{P}_{(hom)}\bigg)^\textit{m} \times \textrm{P}_{(hcorr)}$
\end{center}

where, SC is the set of received chains, $ c_{i}  $ is a chain $ \in $ SC and \textit{i} $ \in $ $ \mathbb{N} $ such that $ 1 \le i \le \sigma $.

\section{Conclusion and Discussions}
\label{lbl:Conclusion}
In this paper, we introduced a new timestamp-free solution to prevent block-withholding or \emph{selfish} mining.  
The key idea of our solution, Zeroblock algorithm, is that each block must be generated and received by the network within \emph{a maximum acceptable time for receiving a new block} interval that is locally estimated by each miner. Within this interval, $mat$, a \emph{honest} miner has to receive or to discover a new block. Otherwise, it generates a dummy block that does not need any proof-of-work computation. The computation of each $mat$ interval is done locally by each miner based on the following Bitcoin parameters: the \emph{expected delay for a block mining} (estimated based on the Poisson nature of the proof-of-work) and the \emph{information propagation time} in the Bitcoin network (that has been previously estimated in \cite{Decker}). 
We prove that our Zeroblock algorithm is resilient to withholding attack. 
That is, we demonstrated that if a \emph{selfish} miner wants to keep its blocks private more than the duration of a $mat$ interval,  then the selfish block will be rejected by the \emph{honest} miners. 
Moreover we prove that selfish miners are not incentivized to ignore dummy Zeroblock or to generate too many of them. 
Furthermore, we demonstrated that our solution is compliant to nodes churn. That is, nodes that freshly enter the system are able to retrieve the \emph{correct chain} provided that a majority of nodes are \emph{honest}. 

Note that Zeroblock scheme is not a solution to \emph{accidental} forks since these forks are not the result of a block-withholding attack. Contrary to \emph{intentional} forks,   \emph{accidental} ones are the result of the Poisson nature of  proof-of-work in Bitcoin. Therefore, with non zero probability two blocks may be discovered by two different pools at almost same times. In Bitcoin network the probability of an \emph{accidental} fork is $\approx 1.69\%$ \cite{Decker}. 
In order to solve this problem an alternative to proof-of-work should be find.
\begin{figure*}
\centering
\makebox[\textwidth][c]{\includegraphics[width=0.5\textwidth]{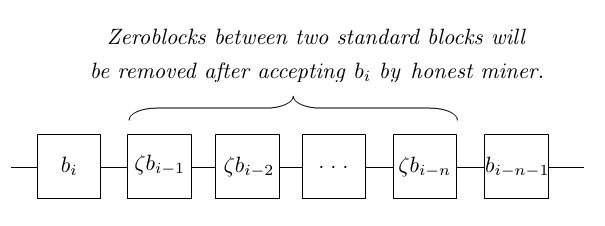}}
\caption{This Figure demonstrates a situation when between standard blocks $b_{i}$ and $b_{i-n-1}$ there are some Zeroblocks from $\zeta b_{i-1}$ to $\zeta b_{i-n}$ that will be removed after accepting standard block $b_{i}$ by honest node. Note that even after removing Zeroblocks from the Blockchain, the standard block $b_{i}$ still includes the hash of Zeroblocks ($\zeta b_{i-1}, ... , \zeta b_{i-n}$) and thus in the future, all of honest nodes are able to understand that between $b_{i}$ and $b_{i-n-1}$ there have been Zeroblocks.} 
\label{fig:ZB_Note(1)}
\end{figure*}

A criticism to our Zeroblock algorithm can be the fact that the blockchain may include too many sequences of dummy Zeroblocks. This can be easily solved as follows.
Each honest miner after receiving and accepting a new standard block $b_{i}$ removes all Zeroblocks between $b_{i}$ and previous standard block (see Figure \ref{fig:ZB_Note(1)}).

ZeroBlock solution is a step further in solving one of the major problems in Bitcoin and can be used also as an \emph{altcoin}, (a term refers to a cryptocurrency based on the Blockchain technology \cite{Sok}) or in conjunction with other cryptocurrencies.

\newpage
\begin{figure*}
\centering
\makebox[\textwidth][c]{\includegraphics[width=1\textwidth]{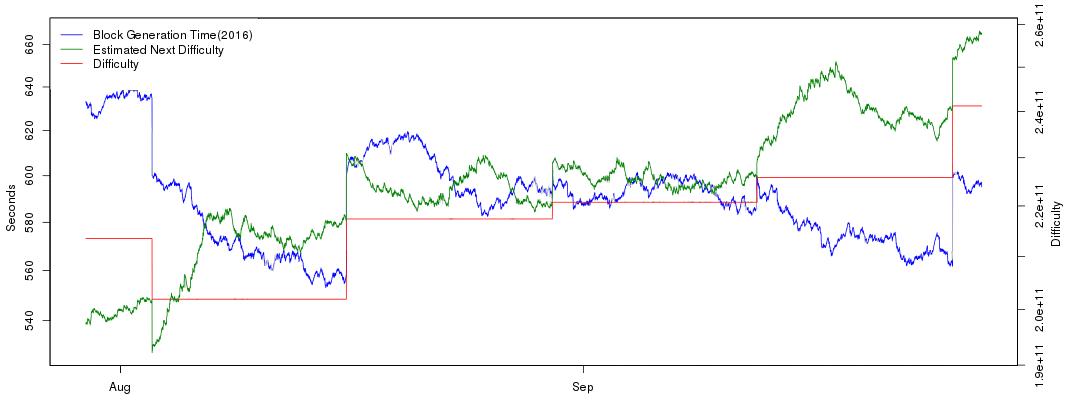}}
\caption{This figure represents the relation between block generation time and difficulty of proof-of work such that using equation \ref{eq6} when 2016 blocks are generated in less than 600 seconds (10 minutes), the difficulty is increased and if they are discovered in more than 10 minutes, then difficulty is decreased.\cite{Bitcoin bitcoinwisdom}}
\label{fig:bitcoin-difficulty_4}
\end{figure*}

\begin{figure*}
	\centering
	\makebox[\textwidth][c]{\includegraphics[width=0.7\textwidth]{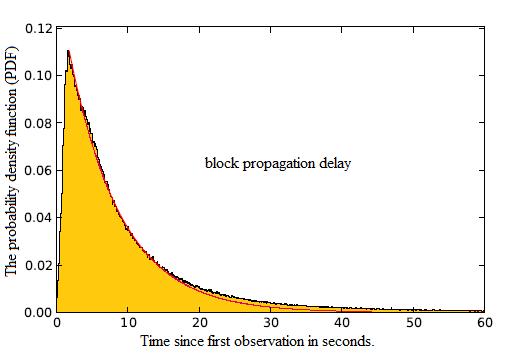}}
	\caption{Decker and Wattenhofer's simulation \cite{Decker} to estimate block propagation delay in the entire Bitcoin network shows that before 60 seconds the whole of network receives a published discovered block. We call this delay \textit{ipt} (\textit{information propagation time})} 
	\label{fig:IPT}
\end{figure*}

\end{document}